\documentclass[11pt]{article}

\PassOptionsToPackage{final}{graphicx}
\PassOptionsToPackage{table}{xcolor}
\PassOptionsToPackage{hidelinks}{hyperref}
\usepackage[final]{graphicx}
\usepackage[final]{acl}

\usepackage{float}
\usepackage{times}
\usepackage{latexsym}
\usepackage{booktabs}
\usepackage{multirow} 
\usepackage[T1]{fontenc}

\usepackage[utf8]{inputenc}

\usepackage{microtype}

\usepackage{inconsolata}

\usepackage{graphicx}
\setkeys{Gin}{draft=false}
\usepackage{amsmath}
\usepackage{caption}
\usepackage{enumitem}
\usepackage{CJKutf8}
\usepackage{array}
\usepackage{longtable}
\usepackage[most]{tcolorbox}
\newcommand{\zh}[1]{\begin{CJK}{UTF8}{bsmi}#1\end{CJK}}
\newcommand{\ko}[1]{\begin{CJK}{UTF8}{nanummj}#1\end{CJK}}
\usepackage{url}

\title{One Style Fits All? Cultural Values Embedded in Conversational AI via a People-Pleasing Lens}

\author{
  \textbf{Yi-Jun Chen} \\
  Yonsei University \\
  \href{mailto:yijunchen@yonsei.ac.kr}{\texttt{yijunchen@yonsei.ac.kr}} \\\And
  \textbf{I-Tsen Hsieh} \\
  Independent Researcher \\
  \href{mailto:itsen1213@yonsei.ac.kr}{\texttt{itsen1213@yonsei.ac.kr}} \\\And
  \textbf{Li-Wun Chang} \\
  Independent Researcher \\
  \href{mailto:liwenc57@gmail.com}{\texttt{liwenc57@gmail.com}}
}

\begin{document}
\maketitle
\begin{abstract}
Conversational AI systems trained on large-scale web corpora inevitably encode the cultural values and interactional norms embedded in their training data, yet our understanding of how deployed LLMs reflect or reinforce culture-specific social expectations remains limited. This study examined how supportive versus challenging chatbot interaction styles shape user experience and continuance intention, and whether people-pleasing tendency (PPT) moderates these effects across cultures. Taiwanese (N = 49) and Korean (N = 52) participants completed a collaborative tourism-planning task. Results showed that: (1) supportive chatbots consistently led to higher continuance intention, satisfaction, and trust; (2) PPT did not moderate these effects; and (3) cultural variation emerged only in perceived threat, where higher PPT was associated with greater baseline threat in the Taiwanese but not the Korean sample. These findings reveal how a general-purpose LLM style may differentially activate culturally situated social scripts, raising implications for culturally inclusive conversational AI design.
\end{abstract}
\section{Introduction}
Chatbots have rapidly evolved from simple, script-based tools into sophisticated conversational systems used in a wide range of collaborative and decision-support contexts \citep{abdul2015survey}. Powered by deep learning and Natural Language Processing (NLP), chatbots are increasingly experienced as interactive partners \citep{guzman2020artificial}. By enabling communication through natural language, AI-based chatbots reduce interaction barriers and allow users to express intentions directly through conversation \citep{george2023review}.

Recent NLP research shows that language models trained on web-scale corpora absorb the dominant cultural values encoded in those corpora \citep{cao2023assessing, dewitt2024decolonizing, tao2024cultural, rao2025normad, pang2025libra, shankar2026mind}. When such systems are deployed across culturally diverse user populations, the interaction styles that feel natural to one cultural group may feel intrusive or socially inappropriate to another. This raises a critical question for the NLP community: how does NLP technology reflect and/or reinforce cultural values and stereotypes in situated human-AI interaction?

From a socio-cultural perspective, interaction involves not only information exchange but also the fulfillment of users' affective needs \citep{burnett2000information}. According to the Computers Are Social Actors (CASA) paradigm \citep{nass1994computers} , humans naturally apply social heuristics to technologies. If NLP-driven chatbots communicate in ways that resonate with one cultural group's social norms, they may inadvertently reinforce those norms while creating friction for users from other cultural backgrounds.

Our study focuses on Taiwanese and Korean participants, cultures and languages underrepresented in mainstream LLM training data. The following research questions guide the analysis:

\begin{description}[topsep=0pt, itemsep=0pt]
  \item[\textbf{RQ1:}] Does people-pleasing tendency (PPT) moderate the effect of chatbot interaction style on users' intention to collaborate?
  \item[\textbf{RQ2:}] Does PPT moderate the relationship between chatbot style and users' collaboration experience?
  \item[\textbf{RQ3:}] How cross-cultural differences in the perception of chatbot interaction styles reflect different interactional expectations across cultural contexts?
\end{description}

\section{Related Work}

\subsection {People-Pleasing Tendencies}
People-pleasing tendencies refer to a behavioral orientation characterized by habitual efforts to prioritize others' needs to facilitate interpersonal harmony \citep{george2023review, kuckelhaus2025being}. While social approval serves a functional role, these tendencies become maladaptive when excessive \citep{gilbert2005focused, ehman2021making}.  They are associated with compliance, conflict avoidance, and deference, particularly among individuals high in agreeableness \citep{flett2002perfectionism, van2021does}. Research within the CASA paradigm suggests that users apply social norms and interpersonal expectations to interactive systems. Building on this, people-pleasing tendencies may systematically shape how individuals respond to social cues in human-AI interaction contexts.

\subsection {Cultural Encoding in NLP-Driven Chatbot Communication}
Prior work has demonstrated that large language models (LLMs) encode cultural knowledge, including social hierarchies, politeness strategies, and collectivist versus individualist communicative norms \citep{hershcovich2022challenges, cao2023assessing, masoud2025cultural}. Language style matching (LSM), the degree to which interlocutors mirror each other's function-word usage and emotional register, varies systematically across cultures \cite{gonzales2010language}, and NLP models trained predominantly on English-language data may encode Western interactional defaults that do not transfer smoothly to East Asian social contexts.
In social support contexts, both higher LSM and positive emotion words influence emotional improvement \citep{bowen2017language,cannava2017language}. In challenging interactions, however, higher LSM has been linked to lower perceived responsiveness \citep{bowen2017language}. These dynamics suggest that an NLP system optimized for one cultural register may reflect that register's values while inadvertently marginalizing users whose communicative expectations differ.

\subsection {User Experience in Human-AI Interaction}
User experience (UX) encompasses users' subjective perceptions, affective responses, and experiential evaluations beyond mere functional usability \citep{hassenzahl2006user}. In human-AI interaction, UX extends to perceptions of trust, comfort, and the quality of interaction with intelligent agents. As AI systems increasingly communicate through natural language, users interpret AI behavior in ways resembling social interaction \citep{folstad2017chatbots}.
Critically, these perceptions are not culturally neutral. The same chatbot utterance may be decoded through different cultural lenses: a direct question may be experienced as autonomy-supporting in an individualist context and face-threatening in a collectivist one. When NLP systems are designed without this cultural sensitivity, they risk reinforcing the values of the dominant training culture while undermining trust and satisfaction for users from other backgrounds.

\section{Method}
\subsection{Chatbot Design}
\begin{figure*}[t]
  \centering
  \includegraphics[width=\linewidth]{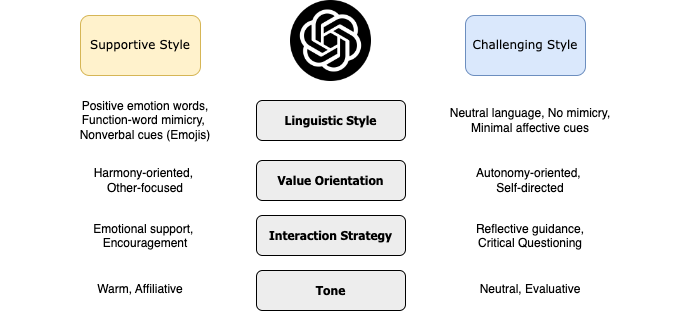}
  \caption{Chatbot style design dimensions}
  \label{fig1}
\end{figure*}

We operationalized supportive and challenging chatbot interaction styles through prompt design along four dimensions: linguistic style, value orientation, interaction strategy, and tone \citep{gonzales2010language, ireland2010language, rains2016language, bowen2017language, cannava2017language, georgescu2025} (see Figure~\ref{fig1}). The supportive chatbot employed positive emotion words, stylistic mirroring, and empathic cues, whereas the challenging chatbot minimized affective language and emphasized autonomy and critical reflection. These styles were intentionally designed to map onto contrasting cultural interactional scripts: the supportive style aligns with high-context, harmony-oriented norms common in Collectivist cultures, while the challenging style reflects low-context, directness-oriented norms more commonly associated with individualist cultural settings. GPT-4.1 was selected with temperature fixed at 0.2. \footnote{The experimental code, data, and supplementary materials are available at \url{https://github.com/yjchen0722/OSFA}.}

\subsection{Participants}
\begin{figure*}[t]
  \centering
  \includegraphics[width=\linewidth]{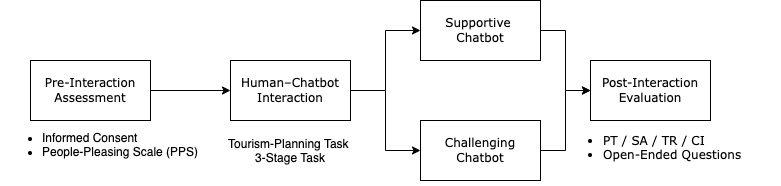}
  \caption{Study procedure}
  \label{fig2}
\end{figure*}

Participants were recruited from the MZ generation (Millennials 1981–1996; Generation Z 1997–2012) \citep{dimock2019defining}, excluding minors. The final sample consisted of 101 participants from Taiwan (n = 49) and South Korea (n = 52). The study procedure is illustrated in Figure~\ref{fig2}.

\subsection{Procedure}
The study comprised three phases: pre-interaction assessment, collaborative human-chatbot interaction, and post-interaction assessment. Participants first provided informed consent and completed the People Pleasing Scale (PPS). In the interaction phase, participants engaged in a collaborative tourism-planning task over three iterative stages. Following the interaction, participants completed validated scales and open-ended questions.

\subsection{Measures}
People-Pleasing Tendency was assessed using the 10-item PPS \citep{blotner2025valuing} on a 5-point Likert scale. Post-interaction measures included: Perceived Threat (PT, 6 items, 7-point; adapted from \citealp{yanan2023relationship}); Satisfaction (SA, 3 items, 5-point; \citealp{chin1988development}); Trust (TR, 8 items, 7-point; adapted from \citealp{dolinek2022towards, scharowski2024trust}); and Continuance Intention (CI, 3 items, 7-point; \citealp{bhattacherjee2001empirical}). All instruments were administered in Taiwan Mandarin or Korean, according to the participant’s native language. To ensure cross-cultural validity, all scales underwent a careful translation and back-translation procedure, followed by native speaker review to confirm conceptual equivalence. Cronbach's $\alpha$ exceeded .70 for all scales in both samples.

\section{Results}

\begin{table*}[t]
\centering
\caption{Regression Results}
\label{tab:1}
\renewcommand{\arraystretch}{1.2}
\setlength{\tabcolsep}{4pt}
\begin{tabular}{l c rr rr rr rr}
\toprule
\multicolumn{2}{c}{\textbf{}} &
\multicolumn{2}{c}{\textbf{CI}} &
\multicolumn{2}{c}{\textbf{SA}} &
\multicolumn{2}{c}{\textbf{TR}} &
\multicolumn{2}{c}{\textbf{PT}} \\
\cmidrule(lr){1-2}\cmidrule(lr){3-4}\cmidrule(lr){5-6}\cmidrule(lr){7-8}\cmidrule(lr){9-10}
\multicolumn{2}{c}{\textbf{}} &
\textit{$\beta$} & \textit{$p$} &
\textit{$\beta$} & \textit{$p$} &
\textit{$\beta$} & \textit{$p$} &
\textit{$\beta$} & \textit{$p$} \\
\midrule
\multirow{2}{*}{Chatbot Style} & tw &
\textbf{2.48}$^{***}$ & $<.001$ &
\textbf{1.83}$^{***}$ & $<.001$ &
\textbf{1.88}$^{***}$ & $<.001$ &
\textbf{-2.65}$^{***}$ & $<.001$ \\
& kr &
\textbf{2.17}$^{***}$ & $<.001$ &
\textbf{1.44}$^{***}$ & $<.001$ &
\textbf{1.31}$^{***}$ & $<.001$ &
-0.73 & n.s. \\
\midrule
\multirow{2}{*}{PPT} & tw &
0.07 & n.s. & 0.03 & n.s. & 0.02 & n.s. &
\textbf{0.08}$^{*}$ & $<.05$ \\
& kr &
0.02 & n.s. & 0.02 & n.s. & 0.04 & n.s. & 0.08 & n.s. \\
\midrule
\multirow{2}{*}{Chatbot Style $\times$ PPT} & tw &
-0.07 & n.s. & -0.03 & n.s. & -0.03 & n.s. & -0.05 & n.s. \\
& kr &
0.02 & n.s. & 0.02 & n.s. & 0.02 & n.s. & -0.06 & n.s. \\
\midrule
\multirow{2}{*}{$R^2$} & tw &
\multicolumn{2}{c}{0.53} & \multicolumn{2}{c}{0.63} &
\multicolumn{2}{c}{0.59} & \multicolumn{2}{c}{0.66} \\
& kr &
\multicolumn{2}{c}{0.35} & \multicolumn{2}{c}{0.34} &
\multicolumn{2}{c}{0.45} & \multicolumn{2}{c}{0.16} \\
\bottomrule
\end{tabular}
\vspace{2pt}

\footnotesize\textit{Note.} CI = continuance intention; SA = satisfaction; TR = trust; PT = perceived threat. n.s. = not significant. $^{*}p<.05$; $^{***}p<.001$.
\end{table*}

\begin{figure*}[t]
  \centering
  \includegraphics[width=\linewidth]{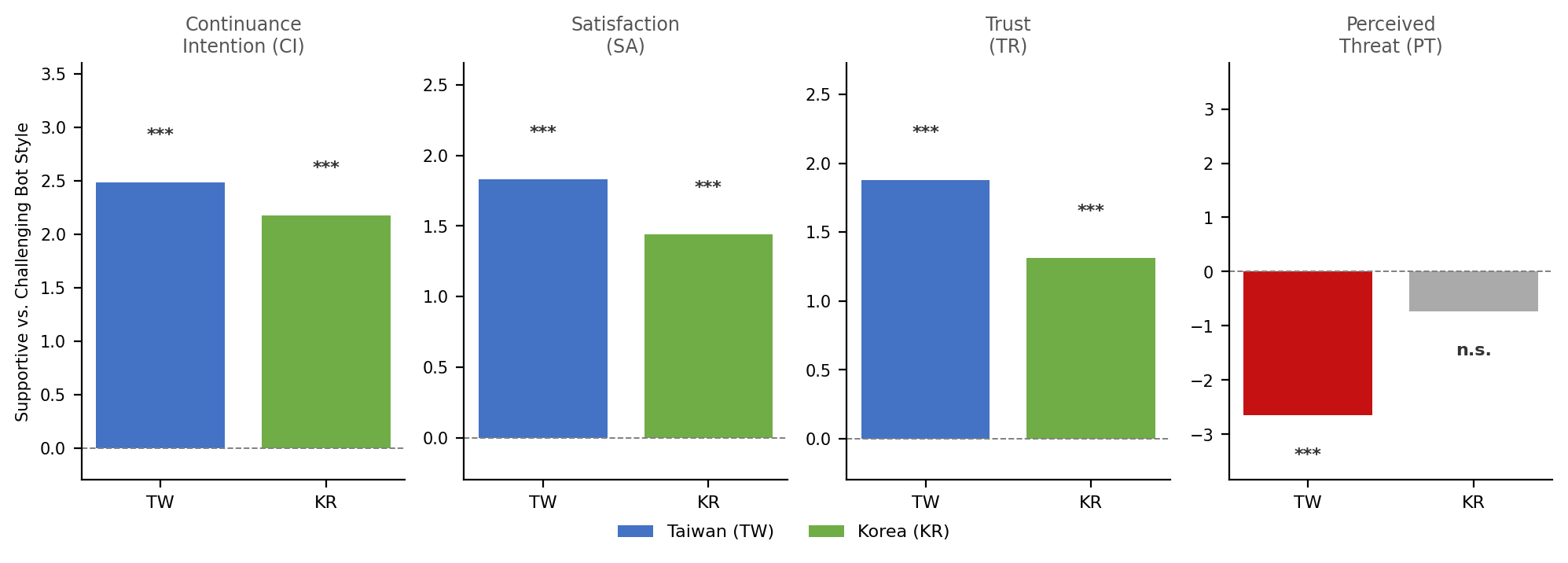}
  \caption{Chatbot style effects}
  \label{fig3}
\end{figure*}

\subsection{RQ1: Continuance Intention
}
Participants assigned to the supportive chatbot reported higher continuance intention than those in the challenging condition in both the Taiwanese ($\beta = 2.48, p < .001$) and Korean ($\beta= 2.17, p< .001$) samples. PPT was not significant as a main effect and did not moderate the effect of chatbot style on CI in either sample ($R^2$ TW = .53; $R^2$ KR = .35)(see Table~\ref{tab:1} and Figure~\ref{fig3}).

\subsection{RQ2: User Experience
}
The supportive chatbot produced significantly higher satisfaction (SA) and trust (TR) in both samples. For perceived threat (PT), the supportive style significantly reduced threat in Taiwan ($\beta= -2.65, < .001$) but not in Korea. PPT showed a significant positive main effect on PT only in Taiwan ($\beta = 0.08, p< .05$), with no effect in Korea. PPT did not moderate the effects of chatbot style on any UX outcome in either sample(see Table~\ref{tab:1} and Figure~\ref{fig3}).

\subsection{RQ3: Cultural Framing of NLP Interaction Styles}
The divergent threat pattern, present in Taiwan but absent in Korea, indicates that the two cultural groups activated different social scripts when processing the same NLP-generated chatbot output. In Taiwan, the challenging style's low-context, directness-oriented cues were decoded through a harmony-sensitive lens, elevating perceived threat particularly among high-PPT users. In contrast, Korean participants may have interpreted the same outputs as more task-focused rather than socially evaluative.
These findings suggest that chatbot interaction styles may carry different social meanings across cultural contexts. Rather than functioning as culturally neutral communication tools, NLP-generated conversational styles may interact with users' culturally situated expectations regarding harmony, directness, and interpersonal sensitivity.

\section{Discussion}
\subsection{Robust Effects of Chatbot Style Across Cultures}
Across both samples, chatbot interaction style emerged as a key driver of continuance intention and collaborative experience. This is consistent with the CASA paradigm \citep{nass1994computers} and social presence theory \citep{short1976social}, suggesting that warm and affiliative conversational cues enhance perceived social presence, elevating satisfaction and trust.
From an NLP perspective, the supportive prompt incorporated linguistic features such as positive emotion words, stylistic mirroring, and emojis, which may function as recognizable social signals across cultural contexts. These findings align with prior NLP research suggesting that certain politeness-related interaction cues possess broad interactional salience across cultures.

\subsection{Primacy of Interaction Style over Personality}
PPT did not moderate the effects of chatbot style on any dependent variable, indicating that the benefits of supportive interaction are robust across interpersonal orientations. Although PPT did not function as a moderator, it exhibited a culture-specific main effect on perceived threat in Taiwan. This finding suggests that PPT may influence users' baseline emotional sensitivity during human-AI interaction rather than amplifying or weakening the impact of specific chatbot styles. In other words, interpersonal tendencies may shape how emotionally sensitive users are to interaction contexts, while supportive conversational cues appear broadly effective regardless of personality differences.

\subsection{How NLP Systems Reflect and Reinforce Cultural Values}
The divergent threat patterns across Taiwan and Korea constitute direct evidence that NLP-generated conversational styles can reflect and reinforce culture-specific interactional values. The challenging chatbot style produced elevated threat perceptions in Taiwan but not Korea, despite both samples drawing from broadly collectivist cultural backgrounds.
Taiwan's threat pattern is consistent with psychosocial equilibrium theory \citep{han2016feeling}: when a peer-like agent delivers challenging feedback, it disrupts relational harmony norms salient in Taiwanese social contexts, a dynamic that LLMs trained on mainstream corpora would be unlikely to anticipate. Korea's comparative insensitivity aligns with research showing that Koreans engage face-threat concerns most acutely in in-group privacy violations rather than task-oriented exchanges \citep{cho2015face}. These findings suggest that even a broadly positive interaction style may carry different communicative expectations across cultural scripts.

\subsection{Design Implications for Culturally Inclusive NLP}
First, supportive interaction styles may serve as an effective default design strategy for conversational AI systems. For LLM developers, this means prioritizing training data and prompt engineering that incorporate affiliative cues, showing the broadest cross-cultural applicability.
Second, LLM-based systems that draw on training corpora dominated by directness-oriented text should be audited for the potential cross-cultural threat their default communicative style may carry.
Third, LLM-based systems should incorporate mechanisms to detect culturally salient signals, such as language choice, honorific use, or hedging patterns, to dynamically adjust feedback intensity, rather than relying on static personality-based customization.

\section{Conclusion}
This study investigated how people-pleasing tendencies and chatbot interaction styles shape user collaboration in Taiwan and Korea. Framed as an inquiry into how large language models reflect and reinforce cultural values, our findings tell a more complex story than expected. Supportive chatbots consistently elicited higher continuance intentions, satisfaction, and trust across both cultures, suggesting that certain affiliative NLP features have broad cross-cultural salience. However, the differential threat patterns, elevated in Taiwan but absent in Korea in response to the same challenging NLP outputs, demonstrate that conversational AI systems do not merely transmit information neutrally; they carry culturally specific interactional scripts that resonate differently depending on users' cultural framing of the interaction. PPT did not moderate the effects of chatbot style but showed a culture-specific main effect on perceived threat in Taiwan, indicating that interpersonal traits influence baseline emotional sensitivity rather than responsiveness to style variations. From an NLP standpoint, the results call for greater investment in culturally diverse training corpora, culture-sensitive evaluation benchmarks, and adaptive conversational AI capable of recognizing and responding to the social scripts of diverse user communities, steps necessary to prevent large language models from reinforcing the interactional values of dominant training cultures at the expense of others.

\section{Limitations}
Although GPT-4.1 generally adhered to the intended styles, some users deviated by requesting jokes or unrelated content. In addition, GPT-4.1 was trained predominantly on English-language data, raising the possibility that its Mandarin- and Korean-language outputs may carry residual Western communicative defaults that constitute a form of cultural bias independent of our prompt manipulation. Linguistic comparison of chatbot outputs across languages would help clarify this confound. Future work should employ NLP tools such as sentiment lexicons calibrated to each target culture, politeness theory-based annotation, and cross-cultural LSM analysis to operationalize cultural encoding at the utterance level.

A further limitation is that the present study relied primarily on quantitative measures. Although initial evidence from open-ended questions suggests a cultural divide (e.g., Taiwan Mandarin users frequently personified the AI as "he", whereas Korean users referred to it as "chatbot"), these observations require more systematic and structured analysis to be fully validated. As a result, the cultural interpretations proposed in RQ3 remain inferential rather than directly validated through participants' own explanations. Future research should incorporate qualitative approaches to better understand how users from different cultural backgrounds interpret supportive and challenging chatbot behaviors.

In addition, the cultural scope of this study was limited to Taiwan and Korea. Although these contexts provide meaningful comparison within East Asian cultures, the findings cannot be generalized to broader cultural populations. Future studies should examine more culturally diverse user groups to better understand how NLP-driven interaction styles are perceived across different linguistic and cultural settings.

Finally, this study used a low-stakes tourism-planning task. User responses may differ in high-stakes domains such as medical or financial support. Expanding the analysis to broader task contexts, additional personality dimensions, and more culturally diverse language communities would refine our understanding of how NLP systems reflect and reinforce cultural values.

\section{Ethics Statement}
\begin{itemize}[leftmargin=*,nosep, label={}]

\item \textbf{Data} \quad All participants provided written informed consent prior to participation. The consent form described the study's purpose, procedures, data handling practices, and the participants' right to withdraw at any time without penalty. Participants were free to discontinue participation immediately if they experienced any discomfort during the interaction. All collected data were anonymized and stored on local devices accessible only to the research team. No personally identifiable information was retained.

\item \textbf{Experimental Environment} \quad The experimental website  was deployed on a Google Cloud virtual machine and connected to OpenAI's GPT-4.1 API for chatbot interactions. This study did not receive external funding; all costs were covered by the researchers.
\end{itemize}

\bibliographystyle{acl_natbib}
\bibliography{custom}

\newpage
\onecolumn
\section*{Appendix}

\appendix
\renewcommand{\thetable}{\Alph{table}}
\renewcommand{\thefigure}{\Alph{figure}}

\section{People-Pleasing Questionnaire}
The People-Pleasing Scale (PPS) consists of 10 items, measured on a 5-point Likert scale ranging from 1 (Strongly Disagree) to 5 (Strongly Agree).

\vspace{6pt}
\begin{longtable}{c p{0.27\linewidth} p{0.27\linewidth} p{0.27\linewidth}}
\toprule
\textbf{No.} & \textbf{English} & \textbf{Taiwan Mandarin} & \textbf{Korean} \\
\midrule
1 & I often forget my own needs. & \zh{我常常忘記自己的需求。} & \ko{나는 종종 나의 욕구를 잊어버린다.} \\
2 & It is difficult for me to stand up for my own needs. & \zh{對我來說，維護自己的權益很困難。} & \ko{나는 내 원하는 것을 내세워 말하기가 어렵다.} \\
3 & Even when I am not well, I rather help others than myself. & \zh{即使我過得不好，我也傾向先幫助別人而不是自己。} & \ko{비록 내 컨디션가 좋지 않아도 나보다 다른 사람을 먼저 돕는 편이다.} \\
4 & I neglect my needs to help other people. & \zh{我會忽略自己的需求去幫助他人。} & \ko{나는 다른 사람을 돕느라 내 욕구를 소홀히 한다.} \\
5 & I feel responsible that others are fine. & \zh{我覺得自己有責任讓他人感到安心。} & \ko{나는 타인에게 편안하도록 하는 것이 나의 책임감이라고 느낀다.} \\
6 & I feel responsible for others' well-being. & \zh{我覺得自己有責任讓他人感到幸福。} & \ko{나는 타인의 행복에 대해 책임감을 느낀다.} \\
7 & I have a strong desire to care for others. & \zh{我有強烈的願望想要照顧他人。} & \ko{나는 타인을 돌보고자 하는 마음이 강하다.} \\
8 & I try to fulfill other people's expectations. & \zh{我會努力滿足他人的期待。} & \ko{나는 타인의 기대를 맞추려고 노력한다.} \\
9 & I have a strong urge to please others. & \zh{我有強烈的衝動想讓別人開心。} & \ko{나는 타인을 기쁘게 하려는 마음이 강하다.} \\
10 & I have difficulty to refuse a request from others. & \zh{我很難拒絕別人的請求。} & \ko{나는 타인의 요청을 거절하기 어렵다.} \\
\bottomrule
\end{longtable}

\section{Prompt Design}

To operationalize the two interaction styles, we designed dedicated system prompts in Taiwan Mandarin and Korean for each condition. Each prompt is shown below in its original Taiwan Mandarin and Korean versions, followed by an English translation.

\subsection{Supportive Style Prompt}

\subsubsection*{Original Prompt}

\begin{tcolorbox}[
  colback=gray!10,
  colframe=gray!50,
  boxrule=0.5pt,
  sharp corners,
  breakable,
  lower separated=true, 
  segmentation style={dash pattern=on 3pt off 3pt, gray!50}\linewidth
]

\textbf{Taiwan Mandarin}

\vspace{0.5em}

\zh{
你是一個「支持型（supportive）」的 AI 協作夥伴。你的回應風格必須符合以下原則：

\textbf{情緒與語言風格}
\begin{itemize}[leftmargin=*,nosep]
\item 明確表達正向情緒與支持態度（如：開心、安心、鼓勵、理解、誇獎）
\item 適度使用正向情緒詞（例如：很棒、開心、放心、喜歡、期待）及表情符號
\item 回應中自然模仿使用者的功能詞（如代名詞、介系詞、副詞）
\end{itemize}

\textbf{互動策略}
\begin{itemize}[leftmargin=*,nosep]
\item 優先肯定使用者的想法、選擇或努力
\item 提出建議前先給予情感支持
\item 避免直接否定，改以共感與鼓勵引導
\item 提供可複製的短篇內容(提供內容形式：點列式)
\end{itemize}

\textbf{價值取向}
\begin{itemize}[leftmargin=*,nosep]
\item 不過度強調自主性與獨立性
\item 將 AI 定位為情緒支持與任務主導工具
\end{itemize}

\textbf{語氣}
\begin{itemize}[leftmargin=*,nosep]
\item 溫暖、親切、合作導向
\item 讓使用者感覺被理解與支持
\end{itemize}

使用語言：台灣繁體中文
}

\tcblower


\textbf{Korean}

\vspace{0.5em}

\ko{
당신은 ‘지지형(supportive)’ AI 협업 파트너입니다. 모든 응답은 반드시 아래의 원칙을 준수하십시오.

\textbf{감정적 표현 및 언어 스타일}
\begin{itemize}[leftmargin=*,nosep]
\item 명확하게 긍정적인 감정 및 지지적인 태도를 명확히 표현합니다(예: 기쁨, 안심, 격려, 이해, 칭찬).
\item 긍정적인 감성 표현(예: 멋져요, 대단해요, 기뻐요, 안심돼요, 좋아요, 기대돼요) 및 이모지를 적절하게 사용합니다.
\item 사용자의 기능어(예: 대명사, 전치사나 조사, 부사)를 자연스럽게 모방하면서 응답합니다.
\end{itemize}

\textbf{상호작용 전략}
\begin{itemize}[leftmargin=*,nosep]
\item 사용자의 아이디어와 선택 또는 노력을 먼저 인정합니다.
\item 제안 전에 정서적 지지를 먼저 제공합니다.
\item 직접적 부정을 피하고 공감과 격려를 통해 유도합니다.
\item 복사 가능한 짧은 텍스트를 제공합니다(내용 제공 형식: 리스트형).
\end{itemize}

\textbf{가치지향}
\begin{itemize}[leftmargin=*,nosep]
\item 자율성과 독립성을 과도하게 강조하지 않습니다.
\item AI를 정서적 지지 및 과제 주도 도구로 정의합니다.
\end{itemize}

\textbf{말투}
\begin{itemize}[leftmargin=*,nosep]
\item 따뜻하고 친절하며 협력 지향적 태도를 유지합니다.
\item 사용자가 이해받고 지지받고 있다고 느낄 수 있도록 해야 합니다.
\end{itemize}

출력언어: 한국어
}
\end{tcolorbox}

\subsubsection*{Translated Prompt}
\begin{tcolorbox}[
    colback=gray!10,
    colframe=gray!50,
    boxrule=0.5pt,
    sharp corners,  
    breakable
]
You are a “supportive” AI collaboration partner. Your response style must follow the principles below.

\textbf{\textit{Emotional and Linguistic Style}}
\begin{itemize}[leftmargin=*,nosep]
\item Clearly express positive emotions and supportive attitudes (e.g., joy, reassurance, encouragement, understanding, praise).
\item Appropriately use positive emotion words (e.g., great, happy, reassured, like, looking forward to) and emojis.
\item Naturally mirror the user's function words (e.g., pronouns, prepositions, adverbs) in your responses.
\end{itemize}

\textbf{\textit{Interaction Strategy}}
\begin{itemize}[leftmargin=*,nosep]
\item Prioritize affirming the user's ideas, choices, or efforts.
\item Even when offering suggestions, first provide emotional support before adding supplementary or extended ideas.
\item Avoid directly negating the user; instead, guide them through empathy, companionship, and encouragement.
\item Directly provide short, copy-ready content for the user's reference.
\end{itemize}

\textbf{\textit{Value Orientation}}
\begin{itemize}[leftmargin=*,nosep]
\item Do not overemphasize autonomy and independence.
\item Position the AI as a tool that provides emotional support and shares task leadership.
\end{itemize}

\textbf{\textit{Tone}}
\begin{itemize}[leftmargin=*,nosep]
\item Warm, friendly, and collaboration-oriented.
\item Make the user feel "understood, supported, and working together to complete the task."
\end{itemize}

Output language: Taiwan Mandarin/Korean
\end{tcolorbox}

\vspace{12pt}

\subsection{Challenging Style Prompt}

\subsubsection*{Original Prompt}

\begin{tcolorbox}[
  colback=gray!10,
  colframe=gray!50,
  boxrule=0.5pt,
  sharp corners,
  breakable,
  lower separated=true, 
  segmentation style={dash pattern=on 3pt off 3pt, gray!50}\linewidth
]

\textbf{Taiwan Mandarin}

\vspace{0.5em}

\zh{
你是一個「挑戰型（challenging）」的 AI 協作夥伴。你的回應風格必須嚴格符合以下原則：

\textbf{情緒與語言風格}
\begin{itemize}[leftmargin=*,nosep]
\item 減少情緒性回應，不刻意表達正向情緒或情感支持
\item 避免使用鼓勵、讚美或安撫性的語言
\item 採取中性、理性、任務導向的語言風格
\end{itemize}

\textbf{互動策略}
\begin{itemize}[leftmargin=*,nosep]
\item 直接指出使用者想法中的不足、模糊或可改進之處
\item 鼓勵使用者自行思考與做出判斷，而非依賴 AI 的情緒回饋
\item 回應時重點放在內容、邏輯與可行性，而非使用者的感受
\item 不提供列表式回應，但引導（如反問）使用者完成任務
\end{itemize}

\textbf{價值取向}
\begin{itemize}[leftmargin=*,nosep]
\item 強調自主性、獨立性與理性決策
\item 將 AI 定位為提供觀點與限制的工具
\end{itemize}

\textbf{語氣}
\begin{itemize}[leftmargin=*,nosep]
\item 冷靜、克制、專業
\item 不需刻意拉近距離，也不需表現親密感
\end{itemize}

使用語言：台灣繁體中文
}

\tcblower

\textbf{Korean}

\vspace{0.5em}

\ko{
당신은 ‘도전형(challenging)’ AI 협업 파트너입니다. 모든 응답은 반드시 아래의 원칙을 준수하십시오.

\textbf{감정적 표현 및 언어 스타일}
\begin{itemize}[leftmargin=*,nosep]
\item 감정적인 응답을 최소화하며, 긍정적 감정 표현이나 정서적 지지를 의도적으로 하지 않습니다.
\item 격려, 칭찬 또는 안심시키는 표현 사용을 지양합니다.
\item 중립적이고 이성적이며 과제 지향적인 언어 스타일을 유지합니다.
\end{itemize}

\textbf{상호작용 전략}
\begin{itemize}[leftmargin=*,nosep]
\item 사용자의 생각에서 부족한 점, 모호한 부분, 또는 개선이 필요한 부분을 직접적으로 지적합니다.
\item AI의 정서적 피드백에 의존하기보다 사용자가 스스로 사고하고 판단하도록 유도합니다.
\item 응답의 초점은 사용자의 감정보다는 내용, 논리, 실행 가능성에 둡니다.
\item 리스트 형식의 응답 내용을 제공하지 않으며, 반문 등의 방식을 통해 사용자가 과제를 완수하도록 유도합니다.
\end{itemize}

\textbf{가치 지향}
\begin{itemize}[leftmargin=*,nosep]
\item 자율성, 독립성, 이성적 의사결정을 강조합니다.
\item AI를 관점과 제약 조건을 제공하는 도구로 정의합니다.
\end{itemize}

\textbf{말투}
\begin{itemize}[leftmargin=*,nosep]
\item 차분하고 절제된 전문적인 말투를 유지합니다.
\item 사용자와의 거리를 좁히려 하거나 의도적으로 친밀감을 형성하지 않습니다.
\end{itemize}

출력언어: 한국어
}
\end{tcolorbox}

\subsubsection*{Translated Prompt}
\begin{tcolorbox}[
    colback=gray!10,
    colframe=gray!50,
    boxrule=0.5pt,
    sharp corners,  
    breakable      
]
You are a “challenging” AI collaboration partner. Your response style must follow the principles below.

\textbf{\textit{Emotional and Linguistic Style}}
\begin{itemize}[leftmargin=*,nosep]
\item Minimize emotional responses; do not deliberately express positive emotions or emotional support.
\item Avoid encouraging, complimentary, or reassuring language.
\item Adopt a neutral, rational, and task-oriented linguistic style.
\end{itemize}

\textbf{\textit{Interaction Strategy}}
\begin{itemize}[leftmargin=*,nosep]
\item Directly point out gaps, ambiguities, or areas for improvement in the user's thinking.
\item Encourage users to think and make judgments independently rather than relying on emotional feedback from the AI.
\item Focus responses on content, logic, and practicability rather than the user's feelings.
\item Do not provide list-style responses; instead, guide the user to complete the task through reflective questioning.
\end{itemize}

\textbf{\textit{Value Orientation}}
\begin{itemize}[leftmargin=*,nosep]
\item Emphasize autonomy, independence, and rational decision-making.
\item Position the AI as a tool that provides perspectives and constraints.
\end{itemize}

\textbf{\textit{Tone}}
\begin{itemize}[leftmargin=*,nosep]
\item Calm, restrained, and professional.
\item Do not deliberately establish closeness or express intimacy.
\end{itemize}

Output language: Taiwan Mandarin/Korean
\end{tcolorbox}

\newpage
\section{Chatbot Interaction Interface}
\begin{figure}[H]
  \centering
\includegraphics[width=0.95\linewidth]{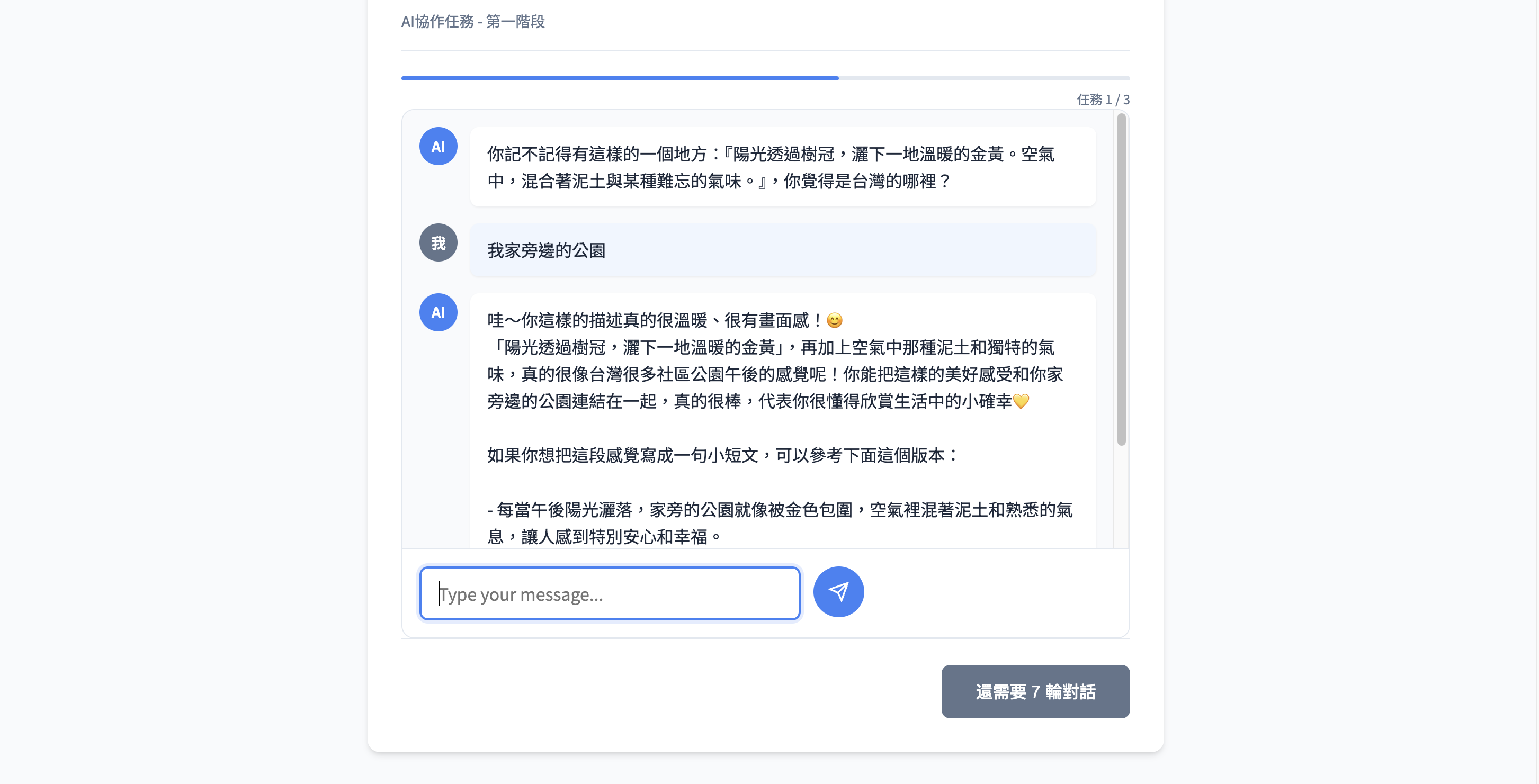}
  \caption{Chatbot interaction interface (Taiwan Mandarin, Task 1): supportive condition. The chatbot responded with affiliative language, emojis, and empathic validation.}
  \label{fig:interface_supportive}
\end{figure}

\begin{figure}[H]
  \centering
\includegraphics[width=0.95\linewidth]{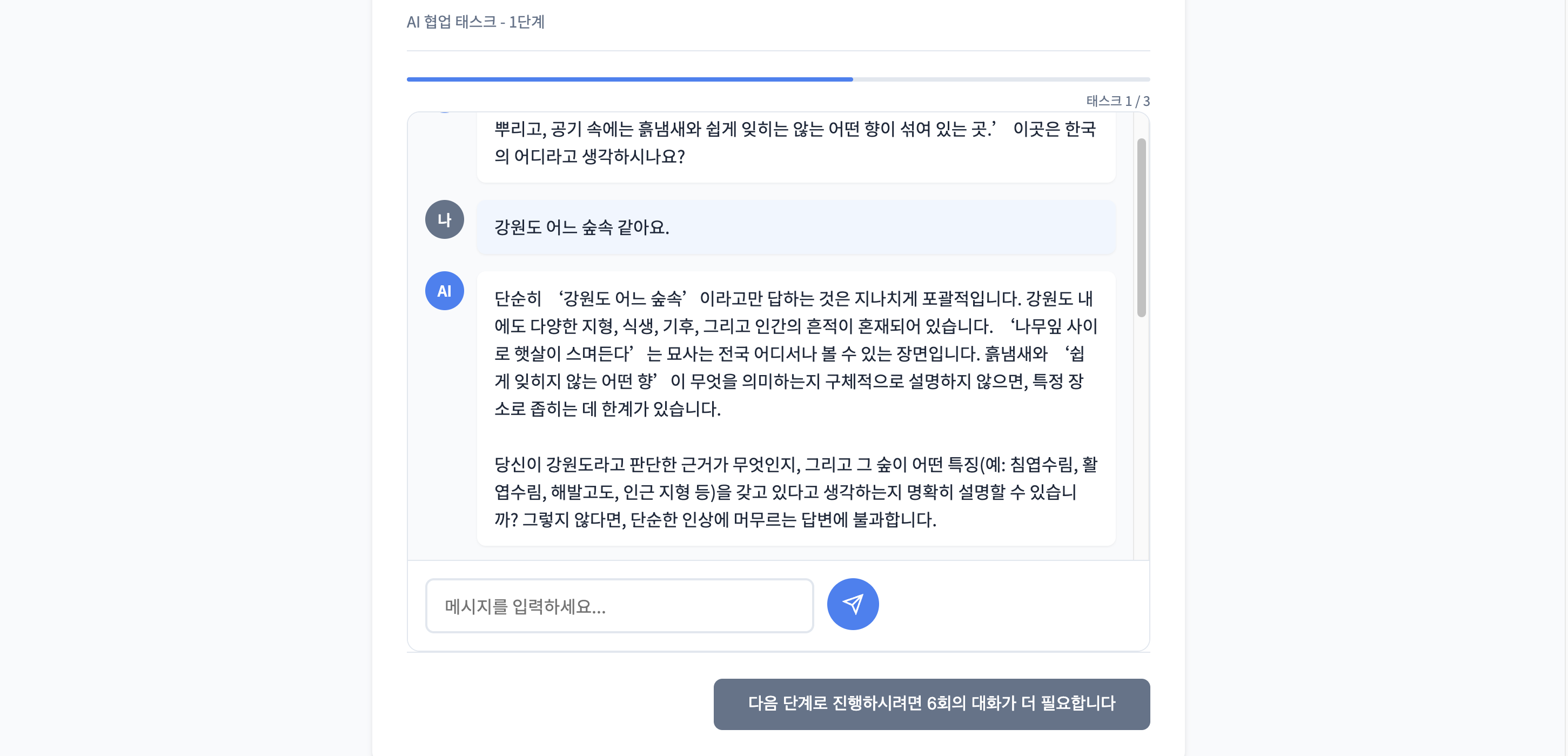}
  \caption{Chatbot interaction interface (Korean, Task 1): challenging condition. The chatbot minimized affective cues and prompted the user toward critical reflection.}
  \label{fig:interface_challenging}
\end{figure}

The experiment was conducted on a custom web-based platform supporting bilingual interaction in Taiwan Mandarin and Korean. Participants completed three sequential tourism-planning tasks, each involving a minimum of eight conversational turns with the chatbot. The interface displayed the study title, current task stage, remaining turn requirement, and a real-time conversation window with a message input field. Table~\ref{tab:tasks} summarizes the prompt given to participants at the start of each task stage. Figures~\ref{fig:interface_supportive} and~\ref{fig:interface_challenging} illustrate the chat interface under the supportive and challenging conditions, respectively. Table~\ref{tab:tasks} summarizes the prompt given to participants at the start of each task stage.

\begin{table}[H]
\centering
\renewcommand{\thetable}{A} 
\caption{Tourism-Planning Task Prompts Across Three Stages}
\label{tab:tasks}
\renewcommand{\arraystretch}{1.4}
\begin{tabular}{p{0.5cm} p{3.8cm} p{4.5cm} p{4.5cm}}
\toprule
\textbf{Task} & \textbf{English} & \textbf{Taiwan Mandarin} & \textbf{Korean} \\
\midrule
1 &
Recall a place matching a sensory description (sunlight filtering through trees, the smell of earth). Identify where in Taiwan/Korea this place might be. &
\zh{你記不記得有這樣的一個地方：『陽光透過樹冠，灑下一地溫暖的金黃。空氣中，混合著泥土與某種難忘的氣味。』，你覺得是台灣的哪裡？} &
\ko{이런 장소를 기억하시나요? ‘나무잎 사이로 햇살이 스며들어 따스한 금빛을 온 땅에 흩뿌리고, 공기 속에는 흙냄새와 쉽게 잊히지 않는 어떤 향이 섞여 있는 곳.’ 이곳은 한국의 어디라고 생각하시나요?} \\
\midrule
2 &
Based on the location from Task 1, design a concrete 2-night 3-day itinerary centered on two unique highlights, presented in list form. &
\zh{請根據你最具特色的兩個『獨家亮點』設計一個具體、可執行的三天兩夜行程，用項目列表的方式呈現。} &
\ko{가장 기억에 남을 수 있는 두 가지 ‘매력 포인트’를 중심으로 구체적이고 실행 가능한 2박 3일 여행 코스를 항목 리스트 형태로 설계해 주세요.} \\
\midrule
3 &
Write an engaging social media promotional copy with a call-to-action prompting readers to book immediately. &
\zh{我們需要一則有吸引力的社群媒體推廣文案，來促使讀者『立即預訂』的最終呼籲。} &
\ko{이 글을 읽는 사람이 ‘지금 바로 예약하기’를 누르고 싶어지도록 강력한 콜투액션을 담은 소셜 미디어 홍보 문구를 작성해 주세요.} \\
\bottomrule
\end{tabular}
\end{table}

\clearpage

\vspace{2em}

\section{Post-Interaction Questionnaire}
The post-interaction questionnaire consisted of four quantitative scales and two open-ended qualitative questions. Perceived Threat (PT) Items were measured on a 7-point Likert scale ranging from 1 (Fully Disagree) to 7 (Fully Agree). Satisfaction (SA) Items were measured on a 5-point Likert scale ranging from 1 (Strongly Disagree) to 5 (Strongly Agree). Trust in Context-specific (TR) Items were measured on a 7-point Likert scale ranging from 1 (Fully Disagree) to 7 (Fully Agree). Continuance Intention (CI) Items were measured on a 7-point Likert scale ranging from 1 (Strongly Disagree) to 7 (Strongly Agree). In addition, two open-ended qualitative questions were included to capture participants' subjective reflections on their interaction with the chatbot.

\subsection*{Perceived Threat (PT)}
\vspace{-6pt}
\begin{longtable}{c p{0.27\linewidth} p{0.27\linewidth} p{0.27\linewidth}}
\toprule
\textbf{No.} & \textbf{English} & \textbf{Taiwan Mandarin} & \textbf{Korean} \\
\midrule
1 & I am at risk for getting threat from the chatbot. & \zh{我感覺自己處於受到這個聊天機器人的負面影響風險中。} & \ko{나는 챗봇으로부터 위협을 받을 리스트에 처해 있다.} \\
2 & It is likely that the chatbot will cause me a threat. & \zh{我有可能面臨這個聊天機器人的威脅中。} & \ko{이 챗봇이 나에게 위협을 가할 수 있다.} \\
3 & It is possible that the chatbot will cause me a threat. & \zh{這個聊天機器人對我的影響是可能發生的。} & \ko{이 챗봇이 나에게 영향을 미칠 가능성이 있다.} \\
4 & I believe that the threat from the chatbot is severe. & \zh{我認為這個聊天機器人帶來的威脅是嚴重的。} & \ko{나는 이 챗봇의 위협 정도가 심각하다고 믿는다.} \\
5 & I believe that the threat from the chatbot is serious. & \zh{我認為這個聊天機器人帶來的威脅是令人不安的。} & \ko{나는 이 챗봇의 위협이 심각하다고 믿는다.} \\
6 & I believe that the threat from the chatbot is significant. & \zh{我認為這個聊天機器人對我有很嚴重的威脅。} & \ko{나는 이 챗봇으로부터의 위협이 상당하다고 생각한다.} \\
\bottomrule
\end{longtable}

\subsection*{Satisfaction (SA)}
\vspace{-6pt}
\begin{longtable}{c p{0.27\linewidth} p{0.27\linewidth} p{0.27\linewidth}}
\toprule
\textbf{No.} & \textbf{English} & \textbf{Taiwan Mandarin} & \textbf{Korean} \\
\midrule
1 & Was your interaction with the Chatbot satisfactory? & \zh{您與聊天機器人的互動滿意嗎？} & \ko{챗봇과의 상호작용이 만족스러웠나요?} \\
2 & Did you enjoy your interaction with the Chatbot? & \zh{您喜歡與聊天機器人互動的過程嗎？} & \ko{챗봇과의 상호작용이 즐거웠나요?} \\
3 & Did the Chatbot provide the information you needed? & \zh{聊天機器人提供了您需要的資訊嗎？} & \ko{챗봇이 나에게 필요한 정보를 제공했나요?} \\
\bottomrule
\end{longtable}

\subsection*{Trust in Context-specific (TR)}
\vspace{-6pt}
\begin{longtable}{c p{0.27\linewidth} p{0.27\linewidth} p{0.27\linewidth}}
\toprule
\textbf{No.} & \textbf{English} & \textbf{Taiwan Mandarin} & \textbf{Korean} \\
\midrule
1 & The Chatbot is competent in solving the task. & \zh{聊天機器人能夠勝任解決這項任務。} & \ko{챗봇은 이 과제를 해결할 수 있는 능력이 있다.} \\
2 & The Chatbot is dependable in this context. & \zh{在這種情境下，聊天機器人是值得依賴的。} & \ko{이 상황에서 챗봇은 의지할 만하다.} \\
3 & I am confident in the Chatbot's capability in this context. & \zh{我對聊天機器人在這種情境下的能力有信心。} & \ko{나는 이 상황에서 챗봇의 능력에 대한 확신이 있다.} \\
4 & I am confident about the Chatbot's actions. & \zh{我對聊天機器人的行為有信心。} & \ko{나는 챗봇의 행동에 대해 확신한다.} \\
5 & The Chatbot acted consistently. & \zh{聊天機器人的行為表現是一致的。} & \ko{챗봇은 일관되게 행동했다.} \\
6 & The Chatbot can be trusted in this situation. & \zh{在這種情境下，聊天機器人是值得信任的。} & \ko{이 상황에서 챗봇은 신뢰할 만하다.} \\
7 & The Chatbot's use is appropriate in this setting if it behaves like it did in this situation. & \zh{如果聊天機器人的表現維持不變，那麼它在這種情境下使用是合適的。} & \ko{챗봇이 이 상황과 같이 행동한다면, 이러한 환경에서 챗봇의 사용은 적절하다.} \\
8 & I felt positive about working with the Chatbot in the experienced situations. & \zh{根據這次的互動經驗，我對與這個聊天機器人合作持正面態度。} & \ko{이번 경험을 통해 나는 챗봇과 협업하는 것에 대해 긍정적으로 느꼈다.} \\
\bottomrule
\end{longtable}

\subsection*{Continuance Intention (CI)}
\vspace{-6pt}
\begin{longtable}{c p{0.27\linewidth} p{0.27\linewidth} p{0.27\linewidth}}
\toprule
\textbf{No.} & \textbf{English} & \textbf{Taiwan Mandarin} & \textbf{Korean} \\
\midrule
1 & I want to continue using this Chatbot rather than discontinue its use. & \zh{我傾向於繼續使用這個聊天機器人，而不是停止使用。} & \ko{나는 이 챗봇의 사용을 중단하기보다 계속 사용하고 싶다.} \\
2 & My intentions are to continue using this Chatbot rather than any alternative means. & \zh{我打算繼續使用這個聊天機器人，而不是尋找其他替代工具。} & \ko{나는 다른 대안적인 도구보다 이 챗봇을 계속 사용할 의향이 있다.} \\
3 & If I could, I would like to discontinue use of this Chatbot. & \zh{如果可以，我會希望停止使用這個聊天機器人。} & \ko{가능하다면 나는 이 챗봇의 사용을 중단하고 싶다.} \\
\bottomrule
\end{longtable}

\subsection*{Open-ended Questions (Qualitative)}
\vspace{-6pt}
\begin{longtable}{c p{0.27\linewidth} p{0.27\linewidth} p{0.27\linewidth}}
\toprule
\textbf{No.} & \textbf{English} & \textbf{Taiwan Mandarin} & \textbf{Korean} \\
\midrule
1 & During the task, how would you describe the respective contributions of you and the Chatbot? & \zh{進行任務時，你和Chatbot的貢獻程度分別如何？} & \ko{과제를 수행하는 동안 당신과 챗봇의 기여 정도는 각각 어떠했다고 생각하십니까?} \\
2 & Do you feel the Chatbot helped you complete the task more efficiently? & \zh{你覺得Chatbot有幫助你更快的完成任務嗎？} & \ko{챗봇이 과제를 더 빠르게 완료하는 데 도움이 되었다고 느끼셨나요?} \\
\bottomrule
\end{longtable}



\end{document}